\documentclass[aps,prl,twocolumn,floats,superscriptaddress,tighten,letterpaper,floatfix]{revtex4-2}
\usepackage{bbm,mathtools}
\usepackage[utf8]{inputenc}
\usepackage{amssymb}
\usepackage{amsbsy}
\usepackage{amsmath}
\usepackage{graphicx}
\usepackage{graphics}
\usepackage{setspace}
\usepackage{array}
\usepackage{color}
\usepackage{xcolor}
\usepackage{fontenc}
\usepackage{textcomp}
\usepackage{rotating}
\usepackage{bm}
\usepackage{braket}
\usepackage[colorlinks=true,linkcolor=black,bookmarksopen=false,urlcolor=blue,citecolor=blue]{hyperref}
\hypersetup{pdfpagemode=UseNone}
\usepackage{chngcntr}
\usepackage{subfigure}

\usepackage{comment}
\usepackage{color}
\usepackage{slashed}
\usepackage[colorlinks=true,linkcolor=black,bookmarksopen=false,urlcolor=blue,citecolor=blue]{hyperref}
\usepackage{amsmath}
\usepackage{enumitem}
\usepackage{revsymb}

\let\normalint\int 
\def\nint{\displaystyle\normalint} 

\newcommand{\calI}{{\cal I}}
\newcommand{\calJ}{{\cal J}}
\newcommand{\calF}{{\cal F}}
\newcommand{\calR}{{\cal R}}

\newcommand{\sfeff}{\mathsf{eff}}

\newcommand{\sfeb}{\mathsf{eb}}

\newcommand{\sfMT}{\mathsf{MT}}
\newcommand{\sfe}{\mathsf{e}}

\newcommand{\sfel}{\mathsf{el}}

\newcommand{\veps}{\varepsilon}

\newcommand{\gbar}{\bar{g}}

\newcommand{\gammael}{\gamma_{\mathsf{el}}}

\newcommand{\mb}{m_{\mathsf{b}}}

\newcommand{\St}{\mathfrak{St}}

\newcommand{\im}{\,\text{Im}\,}

\newcommand{\Te}{T_\mathsf{e}}
\newcommand{\Fano}{\mathsf{F}}
\newcommand{\FanoAdj}{\mathsf{F}_{\mathsf{adj}}}

\newcommand{\shot}{S_{\mathsf{shot}}}
\newcommand{\xbar}{\bar{x}}

\begin{document}

\title{
Suppression of Shot Noise in a Dirty Marginal Fermi Liquid 
}

\author{Tsz Chun Wu}
\affiliation{Department of Physics and Astronomy, Rice University, Houston, Texas 77005, USA}
\author{Matthew S. Foster}
\affiliation{Department of Physics and Astronomy, Rice University, Houston, Texas 77005, USA}
\affiliation{Rice Center for Quantum Materials, Rice University, Houston, Texas 77005, USA}

\date{\today}

\begin{abstract}
We study shot noise in a two-dimensional, disordered marginal Fermi liquid (MFL) driven out of equilibrium. 
We consider electrons with Planckian dissipation on the Fermi surface coupled to quantum-critical bosons in the presence of disorder. 
In the non-interacting and strong electron-boson drag limits, MFL effects disappear and our theory reproduces known results. 
For the case where the bosons remain in equilibrium, inelastic scattering strongly suppresses the noise by dissipating 
the injected energy. Interestingly, we find that MFL effects do play a role in this regime, and give a weak \emph{enhancement}
on top of the otherwise strong suppression. 
Our results suggest that shot noise can be strongly suppressed in quantum-critical systems, and this scenario may 
be relevant to recent measurements in a heavy-fermion strange metal.
\end{abstract}

\maketitle



\textit{Introduction}.---Shot noise in electronic transport measures the correlation of current fluctuations
\begin{equation}
\label{eq:shot_def}
	\shot
	=
	\int dt\,
	\left\langle 
	\delta \hat{I}(t)
	\,
	\delta \hat{I}(0)
	+
	\delta \hat{I}(0)
	\,
	\delta \hat{I}(t)
	\right\rangle
\end{equation}
when the system is driven out-of-equilibrium by a voltage $V$ \cite{NLsM6_Kamenev_CUP_11,shot_Buttiker_review_2000}, 
where $\delta \hat{I}(t) = \hat{I}(t) - \left\langle {I}\right\rangle$ is the current fluctuation and $t$ is time. 
The associated Fano factor $\mathsf{F} = S_{\mathsf{shot}}/ (2\, e\, I)$ ($e$ is electric charge, $I = \left\langle I \right\rangle$ is electric current) offers a unique way to probe the effective charge of carriers in mesoscopic systems, such as superconductor tunneling junctions \cite{shot_Zhou_Natelson_Nat_2019}, fractional quantum Hall devices \cite{shot_FQH_Saminadayar_PRL_1997,shot_FQH_dePicciotto_Nat_1997} and quantum-dot Kondo systems \cite{shot_Kondo_Delattre_NatPhy_2009,shot_Kondo_Zarchin_PRB_2008}.

\begin{figure}[hb!]
\centering
\includegraphics[width=0.98\linewidth]{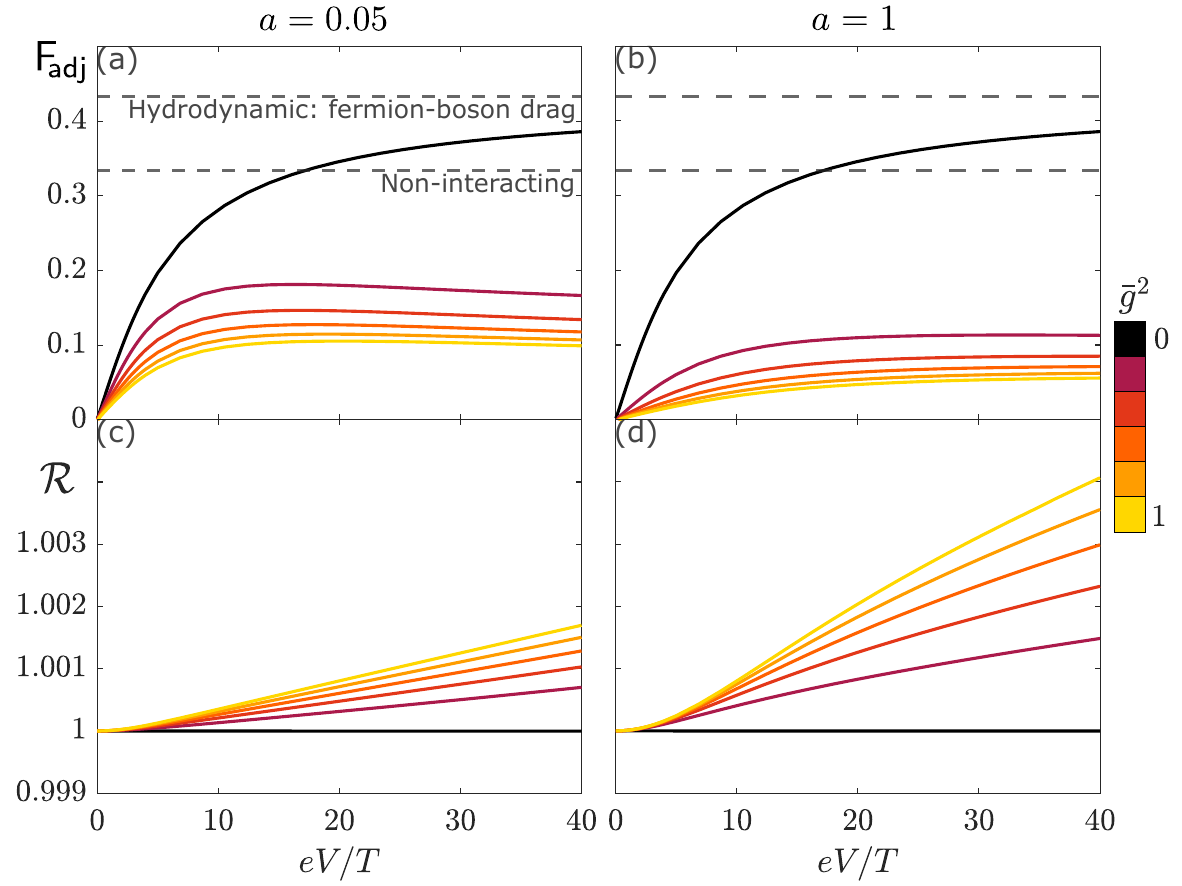}
\caption{
	Plot of the adjusted Fano factor $\FanoAdj$ [Eq.~(\ref{eq:Fano_adj})] (a,b) 
	and 
	the marginal-Fermi-liquid (MFL) correction factor $\calR$  [Eq.~(\ref{eq:calR})] (c,d).
	The latter arises due to the combination of Planckian dissipation (the MFL self-energy) 
	and the Maki-Thompson diagram shown in Fig.~\ref{fig:shot_diagrams}(b).
	All results are plotted as a function of $eV/T$, where $V$ is the voltage and $T$ is the boson temperature, 
	based on Eqs.~(\ref{eq:Fano_finiteT}) and (\ref{eq:calR}) in the equilibrium boson regime. 
	Here 
	the parameter $a \equiv \alpha/\alpha_m$ is inversely proportional 
	to the bosonic thermal mass $\mb^2 = \alpha_m T$ [see Eq.~(\ref{eq:bosonic_GF})]; 
	$\gbar^2$ denotes the reduced squared Yukawa coupling. 
	$\FanoAdj$ is suppressed as the interaction strength $\gbar^2$ increases (black to yellow). 
	The parameters are $\gammael = 10$, $T = 0.05$, and $E_{\mathsf{Th}} = 2\times 10^{-3}$. 
	Our plotted results for $\FanoAdj$ artificially approach $\sqrt{3}/4$ for $\gbar^2 = 0$  and $e V \gg T$ 
	because of the assumed Fermi-Dirac distribution, Eq.~(\ref{eq:F_hydro}). For $\gbar^2 > 0$, the device
	length $L$ is always taken to be much longer than the electron-boson scattering length at the wire center.
	Panels (c,d) indicate that the lack of well-defined quasiparticles gives a \emph{weak enhancement} 
	of the noise, on top of the main effect of dissipative suppression by the boson bath.
}
\label{fig:fano}
\end{figure}

Shot noise in a diffusive wire has been widely studied 
\cite{shot_Buttiker_review_2000,shot_Levitov_JETP_1994,shot_Nagaev_PRB_1995,shot_Nagaev_PhyLettA_1992,shot_Rudin_PRB_1995,shot_Wang_Qi_arxiv_2023,shot_Nazarov_PRL_1994,shot_Nagaev_PhyLettA_1992,Levchenko_Schmalian_review_2020,shot_Beenakker_PRB_1992,NLsM6_Kamenev_CUP_11,shot_Kogan_JETP_1969}. 
In the non-interacting limit, $\Fano = 1/3$ due to the Dorokhov statistics \cite{shot_Levitov_JETP_1994,shot_Nagaev_PRB_1995,shot_Nagaev_PhyLettA_1992,shot_Wang_Qi_arxiv_2023,NLsM6_Kamenev_CUP_11}. 
In the presence of electron-electron interactions, $\mathsf{F}$ acquires a non-universal enhancement at weak coupling \cite{shot_Nagaev_PRB_1995} and becomes universal $\Fano = \sqrt{3}/4$ at $T = 0$ ($T$ is temperature) in the strongly interacting hydrodynamic limit \cite{shot_Rudin_PRB_1995,shot_Nagaev_PRB_1995}. 
Meanwhile, inelastic electron-phonon scattering suppresses $\Fano$ in a non-universal manner 
\cite{shot_Nagaev_PhyLettA_1992,Levchenko_Schmalian_review_2020}. 

In a recent experiment, shot noise measurements on a heavy-fermion strange metal YbRh$_2$Si$_2$ reveal a strongly suppressed Fano factor $\Fano$ \cite{shot_Chen_Natelson_arxiv_2023}. 
Experimental determination of the electron-phonon coupling seemingly rules out electron-phonon scattering as the mechanism \cite{shot_Nagaev_PhyLettA_1992,Levchenko_Schmalian_review_2020}. 
This intriguing result calls for a new theoretical description of shot noise in quantum-critical systems without well-defined quasiparticles.

In this Letter, we investigate the shot noise 
in a two-dimensional (2D), disordered marginal Fermi liquid (MFL) \cite{MFL_Wu_Liao_Foster_PRB_22,disO_SC_MFC_Wu_Lee_Foster_arxiv_2023}.
In this model, electrons are strongly coupled to quantum-relaxational bosons in the presence of quenched disorder, leading to electronic Planckian dissipation.   
Our main physical result is that the \textit{adjusted} Fano factor
\begin{equation}
\label{eq:Fano_adj}
	\FanoAdj(T,V)
	\equiv
	\left(
	S_{\mathsf{shot}}
	-
	S_{\mathsf{JN}}
	\right)
	/
	\left(
	2 \sigma_D e V
	\right),
\end{equation}
which characterizes the excess noise, is strongly suppressed when the quantum-critical bosons remain in equilibrium.
Here $S_{\mathsf{JN}} = 4 T \sigma_D$ denotes the equilibrium Johnson-Nyquist noise
and
$\sigma_{\mathsf{D}}$ is the Drude conductivity.
Throughout this work, we adopt the units $k_B = \hbar = 1$. 

A regime wherein bosons remain in equilibrium can arise 
when interactions amongst critical fluctuations dominate the boson kinetics, instead of boson-fermion scattering.
The corresponding $\FanoAdj$ as a function of $e V / T$ is illustrated in Figs.~\ref{fig:fano}(a,b)
for different values of the dimensionless squared boson-fermion coupling  $\gbar^2 = g^2/(4\pi^2 \gammael)$.
Here $\gammael$ is the impurity scattering rate. 
Excess noise is 
determined 
by the critical interaction through the electron temperature $T_{\sfe}$ [Eq.~(\ref{Tcrossover}), below] 
and the MFL correction to the shot noise. 
The former is 
suppressed 
by the rate of inelastic electron-boson scattering 
\cite{shot_Nagaev_PhyLettA_1992,Levchenko_Schmalian_review_2020}, while the latter 
slightly enhances the Fano factor due to non-equilibrium effects.
We quantify the MFL correction through the ratio $\calR$ illustrated in Figs.~\ref{fig:fano}(c,d). 
The parameter $a \equiv \alpha/\alpha_m$ labeling these plots is inversely proportional to the thermal mass $\mb^2$ of the bosons, see Eq.~(\ref{eq:bosonic_GF}).
Overall, a lighter thermal mass (larger $a$) screens the quantum-critical interaction less and thus boosts the suppression.
Notably, we find that $\FanoAdj$ can fall well below
$1/3$.

\begin{figure}[t!]
\centering
\includegraphics[width=0.65\linewidth]{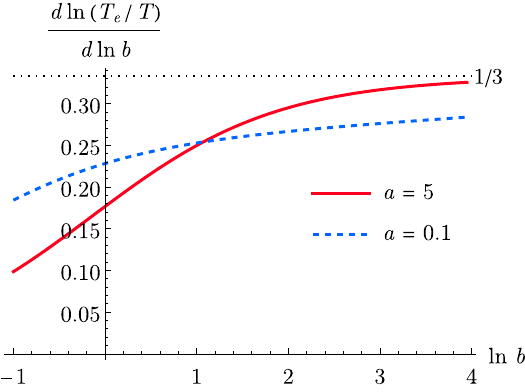}
\caption{Power-law dependence of $T_e/T$ on $b \equiv (e V)^2 E_{\mathsf{Th}}/(8 T^3 \gbar^2)$ for different $a$, see Eqs.~(\ref{Tcrossover}) and (\ref{TcrossoverQuarter}).
}
\label{fig:Teslope}
\end{figure}

\begin{figure}[t!]
\centering
\includegraphics[width=0.9\linewidth]{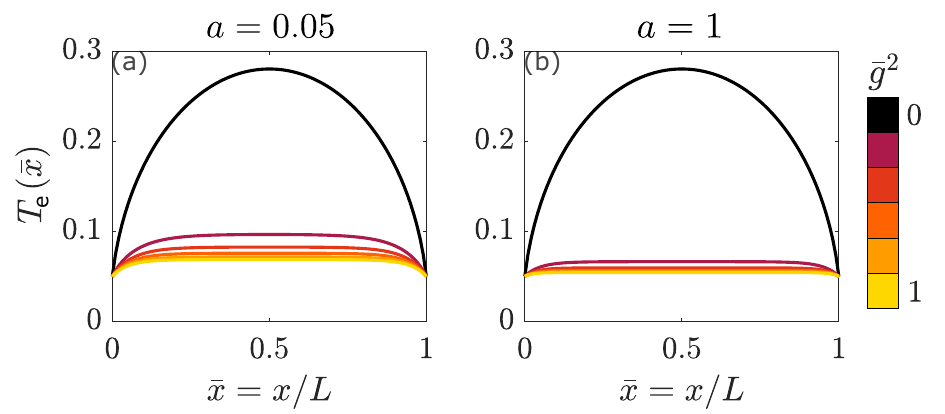}
\caption{
	Plot of the electron temperature profile $\Te(\bar{x})$ obtained by solving Eq.~(\ref{eq:Te_ode}). 
	As the interaction coupling strength $\gbar^2$ increases, $\Te$ decreases, which suppresses the shot noise as a result. 
	With a larger bosonic thermal mass (smaller $a$), the suppression to $\Te$ is weakened due to thermal screening. 
	In the long wire limit with a finite $\gbar^2$, $\Te$ is essentially a constant except in proximity to the contacts. 
	Here, $\gammael = 10$, $T = 0.05$, $E_{\mathsf{Th}} = 2\times 10^{-3}$ and $V = 1$. 
}
\label{fig:Te}
\end{figure}

In the crossover regime where $e V \gtrsim T$, $\FanoAdj$ decreases as $T$ increases [Figs.~\ref{fig:fano}(a,b)], in agreement with the recent 
experimental observation \cite{shot_Chen_Natelson_arxiv_2023}. We find an effective electron temperature that is spatially homogeneous except near the contacts
(see Fig.~\ref{fig:Te}), given by 
\begin{equation}\label{Tcrossover}
	T_{\sfe}
	\simeq
	\left[
		\theta_{\gbar}
		\,
		D (e E)^2
		+
		T^3
	\right]^{1/3},
\end{equation}
where 
$\theta_{\gbar}^{-1} \equiv 2\pi \zeta(3) \bar{g}^2$,
$D$ is the diffusion constant and $E$ is the applied electric field. 
As the interaction strength $\gbar^2$ increases, energy is transferred more rapidly from the electrons to the bosons, which serve as a heat sink, resulting in a more suppressed $\Te$. 
Eq.~(\ref{Tcrossover}) holds for $a \gtrsim 1$. 
For $a \ll 1$, a weaker (roughly $1/4$) 
power-law is instead obtained, 
\begin{equation}\label{TcrossoverQuarter}
	T_{\sfe}(a \ll 1)
	\sim
	\left[
		(c / a \gbar^2)
		\,
		T
		\,
		D (e E)^2
		+
		T^4
	\right]^{1/4}
\end{equation}
where $c$ is a constant, see Fig.~\ref{fig:Teslope}.
Our results in Eqs.~(\ref{Tcrossover}) and (\ref{TcrossoverQuarter}) are reminiscent of a holographic prediction \cite{Sonner2012}.  
The field appears through the combination $(e V)^2 E_{\mathsf{Th}}$ in Eqs.~(\ref{Tcrossover}) and (\ref{TcrossoverQuarter}), relevant in the diffusive regime; here  $E_{\mathsf{Th}} = D/L^2$ is the Thouless energy and
$L$ is the device length. 
The power law in Eq.~(\ref{TcrossoverQuarter}) is consistent with the experimental data in Ref.~\cite{shot_Chen_Natelson_arxiv_2023}, although we do not expect our 2D theory to apply directly to a bulk heavy-fermion material (see \cite{supplemental_material} for a more detailed analysis).
The power in Eq.~(\ref{Tcrossover}) differs from the electron-phonon result \cite{shot_Nagaev_PRB_1995} because the latter depends upon the Debye energy; 
for the 2D MFL, only the dimensionless $\gbar^2$ enters. 

When the bosons and leads reside at $T = 0$, we find 
\begin{equation}
\label{eq:shot_eqm_boson_T=0}
	\FanoAdj(T = 0)
	=
	\Fano
	\simeq
	\frac{2T_{\sfeff}}{eV}
	\,
	\calJ_0
	\left(
	\frac{
	\bar{g}^2
	\pi
	\,
	T_{\sfeff}
	}{\gammael}
	\right),
\end{equation}
where  
$
	\calJ_0(y)
	=
	(1 + y)/(1+y \ln2)
$
and the effective electronic temperature 
$T_{\sfeff} = T_{\sfe}(T = 0)$ in Eq.~(\ref{Tcrossover}).

We focus here on the dirty MFL because recent theoretical studies suggest that disorder is crucial for strange metallicity observed in many quantum materials \cite{SYK_Guo_Patel_linearT_PRB_2022,SYK_Patel_linearT_arxiv_2022,MFL_Wu_Liao_Foster_PRB_22,SYK_Patel_PRB_21}. 
The interplay of disorder, Planckian dissipation, and quantum effects results in interesting consequences for transport \cite{SYK_Guo_Patel_linearT_PRB_2022,SYK_Patel_linearT_arxiv_2022,MFL_Wu_Liao_Foster_PRB_22,SYK_Patel_PRB_21} 
and superconductivity 
\cite{disO_SC_MFC_Sri_Burmistrov_PRB_2023,disO_SC_MFC_Wu_Lee_Foster_arxiv_2023}. 
Due to disorder smearing, the critical bosonic propagator acquires a quantum relaxational form, with the retarded component being \cite{SYK_Guo_Patel_linearT_PRB_2022,SYK_Patel_linearT_arxiv_2022,MFL_Wu_Liao_Foster_PRB_22,Galitski05}
\begin{equation}
\label{eq:bosonic_GF}
	D^R_{\omega,\textbf{q}}
	=
	-
	\left[2(q^2 - i\, \alpha \, \omega\, + \mb^2)\right]^{-1}.
\end{equation}
Here, $q$ is momentum, $\omega$ is frequency,  $\mb^2 = \alpha_m \, T$ is the thermal mass arising due to 
quartic self-interactions amongst the bosons \cite{SYK_Guo_Patel_linearT_PRB_2022,SYK_Patel_linearT_arxiv_2022,MFL_Wu_Liao_Foster_PRB_22,thermal_mass_Torroba}, while $\alpha$ and $\alpha_m$ are model-specific parameters. 
The quantum relaxational  bosons give arise to a MFL self-energy \cite{
MFL_Varma_CuO_PRL_89,
SYK_Patel_PRB_21,
SYK_Guo_Patel_linearT_PRB_2022,
SYK_Patel_linearT_arxiv_2022,
MFL_Wu_Liao_Foster_PRB_22} for electrons at equilibrium.
The fermionic and bosonic propagators are in sharp contrast with their counterparts in the clean limit \cite{NFL_SU_N_Raghu_PRL_19,NFL_SSLee_Review_18,NFL_Subir_book_CUP_2011},
and offer possible sources for the linear-$T$ resistivity observed in strange metals \cite{SYK_Guo_Patel_linearT_PRB_2022,SYK_Patel_linearT_arxiv_2022,MFL_Wu_Liao_Foster_PRB_22,SYK_Patel_PRB_21}. 

\begin{figure}[t]
\centering
\includegraphics[width=0.5\linewidth]{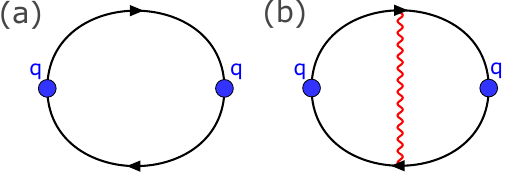}
\caption{
	Feynman diagrams for the Keldysh current-current correlation function: (a) the bubble diagram [Eq.~(\ref{eq:shot_bubble})] and (b) the Maki-Thompson (MT) diagram [Eq.~(\ref{eq:shot_MT})]. 
	Here, the 
	black line 
	is
	the fermion Green's function incorporating
	the MFL self-energy and disorder scattering, while
	the red wavy line represents the quantum relaxational bosonic propagator. 
}
\label{fig:shot_diagrams}
\end{figure}

We investigate the shot noise using the Keldysh formalism \cite{Keldysh_conv4_Kamenev_AdvPhy_09,NLsM6_Kamenev_CUP_11}.
We evaluate the interaction corrections via the bubble [Fig.~\ref{fig:shot_diagrams}(a)] and Maki-Thompson diagrams (MT) [Fig.~\ref{fig:shot_diagrams}(b)] using the non-equilibrium electron distribution function.
The latter is governed by the kinetic equation derived from the MFL-Finkel'stein non-linear sigma model (MFL-FNLsM) \cite{NLsM6_Kamenev_CUP_11,MFL_Wu_Liao_Foster_PRB_22}.  
For bosons staying in equilibrium, we will show that the excess noise is strongly suppressed by the 
dissipation.
On the other hand, in the limit of strong electron-boson drag, contributions from the MFL self-energy and MT diagram cancel and we recover $\Fano = \sqrt{3}/4$ at $T = 0$ \cite{shot_Nagaev_PRB_1995,shot_Rudin_PRB_1995}.



\textit{Model}.---We consider a 2D system with $N$ flavors  of spinless, disordered electrons at finite density coupled to SU($N$) matrix quantum-critical bosons with a Yukawa coupling strength $g$ \cite{MFL_Wu_Liao_Foster_PRB_22,disO_SC_MFC_Wu_Lee_Foster_arxiv_2023,NFL_SU_N_disO_Raghu_PRL_20,NFL_SU_N_Raghu_PRL_19} (formally in the unitary class \cite{10-fold_Altland_Zirnbauer_PRB_1997,10-fold_Ludwig_NewJPhy_2010}).
We restrict our attention to onsite impurity potential disorder and ignore the Sachdev–Ye–Kitaev (SYK)-type randomness in $g$ \cite{SYK_review_Sachdev_RMP_2022}. 

We focus on a wire geometry with a potential difference $V$ applied across the sample along $\hat{x}$. 
We evaluate the shot noise in Eq.~(\ref{eq:shot_def}) by computing the the Keldysh component of the current-current correlator, which describes the correlation of the stochastic fluctuations of the electric current \cite{NLsM6_Kamenev_CUP_11}.

At the semiclassical level, we focus on the contribution due to the bubble [Fig.~\ref{fig:shot_diagrams}(a)] and MT diagrams [Fig.~\ref{fig:shot_diagrams}(b)]:
\begin{equation}
\label{eq:shot=B+MT}
S_{\mathsf{shot}}
=
S_{\mathsf{B}}
+
S_{\mathsf{MT}}
.
\end{equation}
 The bold lines in the diagrams represent the fermionic Green’s function at the saddle-point level \cite{MFL_Wu_Liao_Foster_PRB_22}, encoding the elastic scattering rate due to disorder and the MFL self-energy. 
We ignore the weak-localization \cite{NLsM2_Foster_Liao_Ann_17,NLsM6_Kamenev_CUP_11} and the Altshuler-Aronov quantum corrections \cite{shot_AA_Gefen_PRB_2001,MFL_Wu_Liao_Foster_PRB_22} arising from the interplay between the diffusive collective modes and the quantum-critical bosons.

The shot noise due to the bubble diagram is
\begin{equation}
\label{eq:shot_bubble}
S_{\mathsf{B}}
=
\sigma_{\mathsf{D}}
\int_0^L \frac{dx}{L}
\int
\frac{d\omega}{1 -   \im \Sigma^R_{\omega}(x)/\gammael}
\left[
1 - F_{\omega}^2(x)
\right]
,
\end{equation}
where $\sigma_{\mathsf{D}} = Ne^2 \nu_0 D$ is the Drude conductivity, 
$\nu_0$ is the density of states,
$L$ is the length of the wire,  
and the imaginary part of the non-equilibrium fermionic self-energy is 
\cite{MFL_Wu_Liao_Foster_PRB_22}
\begin{equation}
\label{eq:MFL_selfE_non-eqm}
	\im \Sigma^R_{\omega}(x)
	\!
	=
	\!
	\frac{\gbar^2}{2}
	\int_{-\infty}^{\infty}
	\!\!\!\!\!
	d\nu
	\,
	\tan^{-1}\left( \frac{\alpha \nu}{\mb^2} \right)
	\left[
		F_{\omega + \nu}(x)
		-
		F_{B,\nu}(x)
	\right],
\end{equation} 
where $F_{\veps}(x)$ and $F_{B,\veps}(x)$ are respectively the generalized fermionic and bosonic distribution functions, which are governed by the kinetic equation discussed below. 
The center-of-mass position $x$ dependence arises from the potential gradient applied across the wire. 
In the diffusive limit, the momentum dependence of the distribution functions is very weak and can be ignored. 
The corresponding equilibrium versions are  $F_{\veps}^{\mathsf{eq}} = \tanh(\veps/2T)$ and $F_{B,\veps}^{\mathsf{eq}} = \coth(\veps/2T)$ \cite{Keldysh_conv4_Kamenev_AdvPhy_09,NLsM6_Kamenev_CUP_11,NLsM2_Foster_Liao_Ann_17,MFL_Wu_Liao_Foster_PRB_22}. 
In this case, Eq.~(\ref{eq:MFL_selfE_non-eqm}) reduces to the MFL form \cite{MFL_Wu_Liao_Foster_PRB_22}. 

On the other hand, the contribution from the MT diagram is \cite{supplemental_material}
\begin{equation}
\label{eq:shot_MT}
S_{\mathsf{MT}}
=
\frac{i\, N \,e^2}{2d}
\int_0^L \frac{dx}{L}
I_{\mathsf{MT}}(x)
,
\end{equation}
where $d = 2$ is the dimension, and
\begin{equation}
\label{eq:I_MT}
\begin{aligned}
	I_{\mathsf{MT}}(x)
	\simeq&\,
	2i \,g^2 \, v_F^2
	\int\limits_{\textbf{k},\textbf{q},\omega,\nu}
	\,
	2\,\im D^R_{-\nu,-\textbf{q}}
\\
	&\times
	G^R_{\omega,\textbf{k}}
	G^A_{\omega,\textbf{k}}
	G^R_{\omega + \nu,\textbf{k} + \textbf{q}}
	G^A_{\omega + \nu,\textbf{k} + \textbf{q}}
	\,
	\calF (\omega,\nu)
,
\end{aligned}
\end{equation}
the thermal factor
\begin{equation}
\label{eq:MT_thermal_F}
\begin{aligned}
	\calF (\omega,\nu)
	=&\,
	F_{B,-\nu}
	\left(
	F_{\omega} F_{\omega + \nu} 
	-
	 F_{\omega + \nu }^2
	+
	1
	-
	 F_{\omega}^2
	\right)
\\
	&+
	F_{\omega } F_{\omega + \nu}
	\left(
	F_{\omega + \nu}
	- F_{\omega}
	\right),
\end{aligned}
\end{equation}
and $G^{R/A}_{\omega,\textbf{k}}$ is the retarded/advanced fermionic propagator. 
In the above expression, we have dropped terms that vanish by causality and ignored finite angle scattering terms. 
We explicitly check that $S_{\mathsf{B}}$ and $S_{\sfMT}$ reduce to the Johnson–Nyquist noise at equilibrium \cite{supplemental_material}, as expected by the fluctuation-dissipation theorem \cite{NLsM6_Kamenev_CUP_11,shot_Buttiker_review_2000}. 

It now remains to determine the distribution functions in order to compute the interaction corrections to the shot noise. 
The kinetic equation governing $F_{\omega}$ in the diffusive regime can be derived based on the saddle point of the MFL-FNLsM \cite{supplemental_material,MFL_Wu_Liao_Foster_PRB_22,NLsM6_Kamenev_CUP_11}. 
The result is
\begin{equation}
\label{eq:kinetic_eqn}
-D \, \nabla_x^2 F_{\omega}(x) = \St_{\sfeb}[F],
\end{equation}
where $D = v_F^2/4\gammael$ is the diffusion constant and the electron-boson collision integral is given by
\begin{equation}
\label{eq:St_coll}
\begin{aligned}
&
\St_{\sfeb}[F]
=
4g^2
\int\limits_{\Omega,\textbf{q}}
\im G^{R}_{\omega + \Omega,\textbf{p} + \textbf{q}}
\;
\im D^{R}_{\Omega,\textbf{q}}
\\
&\times
\begin{Bmatrix}
F_{B,\Omega}(x)
[
F_{\omega + \Omega}(x)
-
F_{\omega}(x)
]
-
[
1 
-
F_{\omega + \Omega}(x)
F_{\omega}(x)
]
\end{Bmatrix}
.
\end{aligned}
\end{equation}
In equilibrium, $\St_{\sfeb}[F] = 0$. 
$F_{\omega}(x)$ satisfies the boundary conditions
\begin{equation}
	F_{\omega}(0)
	=
	\tanh\left(\frac{\omega}{2T}\right),
\,\,
	F_{\omega}(L)
	=
	\tanh\left(\frac{\omega - e \,V}{2T}\right).
\end{equation}



\textit{Results}.---In the absence of interactions, 
$\St_{\sfeb}[F] = \im \Sigma^R_{\omega}(x) = 0$ and 
Eq.~(\ref{eq:kinetic_eqn}) has the standard solution 
\begin{equation}
\label{eq:F_NI}
	F_{\omega}^{\mathsf{NI}}(x)
	=
	\left(
	1 - \xbar
	\right)
	\tanh\left(\frac{\omega}{2T}\right)
	+
	\xbar
	\tanh\left(\frac{\omega - e\, V}{2T}\right),
\end{equation}
where $\xbar = x/L$. 
Plugging Eq.~(\ref{eq:F_NI}) into Eq.~(\ref{eq:shot_bubble}) yields the Fano factor $\mathsf{F} = 1/3$.  

With interactions, we focus on the kinetic regime $l_{\sfel} \ll l_{\sfeb} \ll L$, 
where $l_{\sfel}$ ($l_{\sfeb}$) is the elastic (inelastic) scattering length due to impurities (electron-boson collisions).
In this ``hydrodynamic'' limit, both the fermion and boson distribution functions are expected to reflect local equilibration throughout the bulk of the wire, although deviations can occur near the contacts. Even with this constraint, however, different limits are possible, depending upon the relationship between the local electron and boson temperatures.   

If electron-boson collisions occur much more frequently than boson-boson scattering, then 
the bosons are driven away from equilibrium with the electrons and share a common local temperature
(electron-boson drag regime).
The distribution function $F_{\omega}$ acquires a local Fermi-Dirac form
\begin{equation}
\label{eq:F_hydro}
	F_{\omega}^{\mathsf{hydro}}(x)
	=
	\tanh 
	\left[
	\frac{
		\omega - e \, V \, \xbar
	}{
		2T_{\mathsf{e}}(\xbar)
	}
	\right]
\end{equation}
and $F_{B,\Omega}$ acquires a local Bose form 
\begin{equation}
	F_{B,\omega}^{\mathsf{hydro}}(x)
	=
	\coth 
	\left[
	\frac{
		\omega
	}{
		2T_{\mathsf{e}}(\xbar)
	}\right],
\end{equation}
nullifying the collision integral in Eq.~(\ref{eq:St_coll}). 
The local temperature profile $T_{\mathsf{e}}(\xbar)$ can be determined using Eq.~(\ref{eq:kinetic_eqn}) subjected to the boundary conditions $T_{\mathsf{e}}(0) = T_{\mathsf{e}}(1) = T$ \cite{shot_Nagaev_PRB_1995}. 
In this limit, the self-energy term $\Sigma^R_{\omega}(x)$  in $S_{\mathsf{B}}$ [Eq.~(\ref{eq:shot_bubble})] is canceled by $S_{\mathsf{MT}}$ [Eq.~(\ref{eq:shot_MT})]
\cite{supplemental_material},
as also occurs in the dc conductivity calculation \cite{SYK_Guo_Patel_linearT_PRB_2022,SYK_Patel_linearT_arxiv_2022,MFL_Wu_Liao_Foster_PRB_22}.
As a result, $\Fano = \sqrt{3}/4$ for $T \rightarrow 0$ \cite{shot_Nagaev_PRB_1995}.

The physics is more interesting if we assume that the fermions acquire a hydrodynamic distribution function [Eq.~(\ref{eq:F_hydro})] while the bosons remain at equilibrium, i.e. $F_{B,\omega} = F_{B,\omega}^{\mathsf{eq}}$, which can happen due to the boson-boson self-interactions, boson-impurity collisions, and other sources of scattering. 
In this regime, we can integrate both sides of Eq.~(\ref{eq:kinetic_eqn}) to obtain a diffusion equation for the electron temperature profile $\Te$ \cite{shot_Nagaev_PRB_1995}:
\begin{equation}
\label{eq:Te_ode}
	\frac{\pi^2}{6}
	\nabla_{\bar{x}}^2 
	\Te^2(\bar{x})
	=
	-
	(e V)^2
	+
	\frac{1}{2E_{\mathsf{Th}}}
	\int_{-\infty}^{\infty} 
	d\omega
	\,
	\omega
	\,
	\St_{\sfeb}[F],
\end{equation}
where $E_{\mathsf{Th}} = D/L^2$ is the Thouless energy. 
We numerically solve for $\Te$ using Eq.~(\ref{eq:Te_ode}) and the results are shown in Fig.~\ref{fig:Te}.
In the long-wire limit ($L\rightarrow \infty$), the Laplacian term in Eq.~(\ref{eq:Te_ode}) can be dismissed and $\Te$ remains constant 
except in proximity to the contacts. 
At low $T$, $\Te$ can be solved approximately to give Eq.~(\ref{Tcrossover}) \cite{supplemental_material},
which is consistent with the numerics. 

Evaluating the shot noise using Eqs.~(\ref{eq:shot=B+MT}),~(\ref{eq:shot_bubble}),~(\ref{eq:shot_MT}) and~(\ref{eq:F_hydro}), we have 
\begin{equation}
S_{\mathsf{shot}}
=
4
\,
\sigma_{\mathsf{D}}
\,
\int_0^1 d\xbar
\,
\Te(\xbar)
\,
{\cal J}(\xbar,\gbar^2, T)
,
\end{equation}
where
\begin{eqnarray}
\label{eq:calJ}
	{\cal J}(\xbar,\bar{g}^2, T)
	&=&
	\frac{1}{2}
	\int_{-\infty}^{\infty} 
	d \bar{\omega}
	\,
	\frac{
		\calI_{\mathsf{B}}(\xbar,\bar{\omega})
		+
		\calI_{\mathsf{MT}}(\xbar,\bar{\omega})
	}{
		1 - \im \Sigma^R_{2\Te \bar{\omega}}(\xbar)/\gammael
	}
	,
	\\
	\calI_{\mathsf{B}}(\xbar,\bar{\omega})
	&=&
	1 - \tanh^2
	\left(
		\bar{\omega}
		-
		e \, V \, \xbar/2\Te
	\right),
\end{eqnarray}
and
\begin{align}
	\calI_{\mathsf{MT}}(\bar{x},\bar{\omega})
	=
	-
	\frac{
		\bar{g}^2 \Te
	}{
		\gammael
	}
	\nint_{-\infty}^{\infty} 
	\!\!\!\!	
	d \bar{\nu}
	&\,
	\tan^{-1}\left(\frac{2a \, \Te \bar{\nu}}{T}\right)
\nonumber\\
	&\,
	\times
	\calF(2\Te \bar{\omega},2\Te \bar{\nu})
	.
\end{align}
Here, the parameter $a = \alpha/\alpha_m$ and the position dependence on $\Te$ is implicit. 
The corresponding formal Fano factor (defined here for general $e V/T$) 
is 
\begin{equation}
\label{eq:Fano_finiteT}
	\Fano
	=
	\frac{2 }{e \,V}
	\int_0^1 d\bar{x}
	\,
	\Te(\bar{x})
	\,
	{\cal J}(\bar{x},\bar{g}^2, T)
.
\end{equation}
In the integrand, $\Te(\bar{x})$ arises from the solution to the kinetic equation, as occurs for electron-phonon scattering \cite{Levchenko_Schmalian_review_2020}, while $\calJ(\bar{x},\bar{g}^2, T)$ [Eq.~(\ref{eq:calJ})] encodes MFL corrections that slightly enhance $\Fano$. 
Despite its complicated form, $\shot$ can be evaluated analytically at $T = 0$, resulting in Eq.~(\ref{eq:shot_eqm_boson_T=0}) \cite{supplemental_material}.  
Numerical results for $T > 0$ are shown in Figs.~\ref{fig:fano}(a,b).

We quantify the effects of the MFL self-energy and the Maki-Thompson (MT) diagram through the ratio
\begin{equation}
\label{eq:calR}
\!\!\!\!
	{\cal R}(\gbar,T)
	=
	\nint_0^1 d\xbar
	\,
	T_{e}(\xbar)
	\,
	{\cal J}(\xbar,T,\bar{g}^2)
\left[
	\nint_0^1 d\xbar
	\,
	T_{e}(\xbar)
\right]^{-1}
\!\!\!\!\!,
\end{equation}
which is illustrated in Figs.~\ref{fig:fano}(c,d).
Since the bosons and fermions do not share the same temperature, i.e. $T \neq \Te(0 < \bar{x} < 1)$, 
the contribution from $\Sigma^R_{\omega}(x)$ is not canceled, in contrast with the boson-fermion drag and linear response regimes. 
Similar to the phonon suppression mechanism, $\FanoAdj$ decreases as the wire length $L$ and the coupling strength $\gbar^2$ increase. 
The MFL effects encoded in $\calJ$ and ${\cal R}(\gbar,T)$ produce a weak enhancement to the noise.



\textit{Conclusion}.---We have developed a theory for the suppression of shot noise in a 2D disordered MFL 
at the semiclassical level using the Keldysh framework. 
Our theory reproduces known results in the non-interacting and electron-boson drag limits. 
If the quantum-relaxational bosons remain in equilibrium, 
shot noise is strongly suppressed
in the non-universal manner described by Eq.~(\ref{eq:Fano_finiteT}). 
This scenario
may be relevant to the recent experiment \cite{shot_Chen_Natelson_arxiv_2023,supplemental_material}. 

Our theory features a power-law temperature dependence in the shot noise [Eqs.~(\ref{Tcrossover}) and (\ref{TcrossoverQuarter})], 
with an exponent that can differentiate noise suppression mechanisms due to quantum-critical bosons versus phonons \cite{Levchenko_Schmalian_review_2020, shot_Nagaev_PRB_1995}. 
Our results imply that the suppression of shot noise does not necessarily correlate to the absence of quasiparticles \cite{shot_Chen_Natelson_arxiv_2023}. 
In fact, we find that the non-equilibrium MFL self-energy and vertex correction instead \emph{weakly boost} the shot noise on top of the dissipative suppression 
due to the bosons [Figs.~\ref{fig:fano}(c,d)].

Our results contrast with a recent theory proposed in Ref.~\cite{shot_Sachdev_arxiv_2023}, where a universal Fano factor of $1/6$ is obtained in the large-interaction variance limit $g'^{2} \rightarrow \infty$ using the SYK-type model with spatially random Yukawa couplings at $T = 0$. 
The differences 
arise from the form of the electron distribution function used; 
both calculations consider bosons in equilibrium.
We focus on
the kinetic regime where $l_{\sfel} \ll l_{\sfeb} \ll L$, which generically admits electron and boson distribution functions of local equilibrium form \cite{Levchenko_Schmalian_review_2020}. 
Moreover, for a fixed voltage $V$, the 
Fano factor
in Ref.~\cite{shot_Sachdev_arxiv_2023} is independent of $L$ whereas our result vanishes as $L \rightarrow \infty$
(for either fixed voltage $V$ or electric field strength $V/L$). 
Physically, as the length of the nanowire increases, we expect the shot noise to be further suppressed due to 
additional dissipation, consistent with the case of electron-phonon interactions \cite{Levchenko_Schmalian_review_2020}.  

There are numerous intriguing avenues that merit further investigations. 
From a theoretical perspective, it would be interesting to explore the weak-localization 
\cite{NLsM2_Foster_Liao_Ann_17,NLsM6_Kamenev_CUP_11} and Altshuler-Aronov \cite{shot_AA_Gefen_PRB_2001,MFL_Wu_Liao_Foster_PRB_22} quantum corrections to shot noise. 
It would also be valuable to explore the effects of other impurity distributions and the consequences a non-equilibrium bosonic temperature by solving a separate kinetic equation for the bosons. 
On the experimental front, it would be captivating to measure the shot noise in other quantum-critical systems to determine the universality of the results and to test the predicted 
power-law temperature dependence.  On the other hand, it is worthwhile to explore the local temperature profile of the nanowire, which can offer valuable insights into unveiling the physical distribution function.

\textit{Acknowledgments}.---We thank Maxim Dzero, Alex Kamenev, Alex Levchenko, Doug Natelson, and Yiming Wang for useful discussions. 
We are grateful to Doug Natelson and Liyang Chen for sending us the experimental data.
This work was supported by the Welch Foundation Grant No.~C-1809 (T.C.W. and M.S.F.).


\begin{thebibliography}{99}


\bibitem{NLsM6_Kamenev_CUP_11}
	A. Kamenev,
	\textit{Field Theory of Non-Equilibrium Systems}, 2nd ed.
	(Cambridge University Press, Cambridge, England, 2023).
\bibitem{shot_Buttiker_review_2000}
	Ya. M. Blanter, M. Buttiker,
	Shot Noise in Mesoscopic Conductors,
	Phys. Rep. \textbf{336}, 1 (2000).
\bibitem{shot_Zhou_Natelson_Nat_2019}
	P. Zhou, L. Chen, Y. Liu, I. Sochnikov, A. T. Bollinger, M.-G. Han, Y. Zhu, X. He, I. Bozovic, D. Natelson, 
	Electron pairing in the pseudogap state revealed by shot noise in copper oxide junctions, 
	Nature \textbf{572}, 493-496 (2019).
\bibitem{shot_FQH_dePicciotto_Nat_1997}
	R. de-Picciotto, M. Reznikov, M. Heiblum, V. Umansky, G. Bunin, and D. Mahalu, 
	Direct observation of a fractional charge,
	Nature \textbf{389}, 162-164  (1997). 
\bibitem{shot_FQH_Saminadayar_PRL_1997}
	L. Saminadayar, D. C. Glatti, Y. Jin, B. Etienne, 
	Observation of the $e/3$ fractionally charged Laughlin quasiparticle, 
	Phys. Rev. Lett. \textbf{97}, 2526-2529 (1997).
\bibitem{shot_Kondo_Delattre_NatPhy_2009}
	T. Delattre, C. Feuillet-Palma, L. G. Herrmann, P. Morfin, J.-M. Berroir, G. Fève, B. Plaçais, D. C. Glattli, M.-S. Choi, C. Mora, and T. Kontos, 
	Noisy Kondo impurities, 
	Nat. Phys. \textbf{5}, 208 (2009).
\bibitem{shot_Kondo_Zarchin_PRB_2008}
	O. Zarchin, M. Zaffalon, M. Heiblum, D. Mahalu, V. Umansky, Two-electron bunching in transport through a quantum dot induced by Kondo correlations, 
	Phys. Rev. B \textbf{77}, 241303 (2008).
\bibitem{shot_Kogan_JETP_1969}
	Sh. M. Kogan and A. Ya. Shul’man, 
	Theory of Fluctuations in a Noneqilibrium Electron Gas, 
	Zh. Eksp. Teor. ´Fiz. \textbf{56}, 862 (1969) [Sov. Phys. JETP \textbf{29}, 467 (1969)].
\bibitem{shot_Beenakker_PRB_1992}
	C. W. J. Beenakker, M. Büttiker, 
	Suppression of shot noise in metallic diffusive conductors, 
	Phys. Rev. B \textbf{46}, 1889-1892 (1992).
\bibitem{shot_Nagaev_PhyLettA_1992}
	K. E. Nagaev, 
	On the shot noise in dirty metal contacts, 
	Phys. Lett. A \textbf{169}, 103-107 (1992).
\bibitem{shot_Levitov_JETP_1994}
	B. L. Altshuler, L. S. Levitov, A. Y. Yakovets, 
	Nonequilibrium noise in a mesoscopic conductor: A microscopic analysis, 
	JETP Lett. \textbf{59}, 857-863 (1994).
\bibitem{shot_Nazarov_PRL_1994}
	Y. V. Nazarov, 
	Limits of universality in disordered conductors, 
	Phys. Rev. Lett. \textbf{73}, 134-137 (1994).
\bibitem{shot_Rudin_PRB_1995}
	V. I. Kozub and A. M. Rudin, 
	Shot noise in mesoscopic diffusive conductors in the limit of strong electron-electron scattering,
	Phys. Rev. B \textbf{52}, 7853-7856 (1995).
\bibitem{shot_Nagaev_PRB_1995}
	K. E. Nagaev, 
	Influence of electron-electron scattering on shot noise in diffusive contacts, 
	Phys. Rev. B \textbf{52}, 4740-4743 (1995).
\bibitem{shot_Wang_Qi_arxiv_2023}
	Y. Wang, C. Setty, S. Sur, L. Chen, S. Paschen, D. Natelson, Q. Si,
	Shot noise as a characterization of strongly correlated metals, 
	arxiv:2211.11735. 
\bibitem{Levchenko_Schmalian_review_2020}
	A. Levchenko, J. Schmalian,
	Transport properties of strongly coupled electron-phonon liquids,
	Annals of Physics \textbf{419}, 168218 (2020).
\bibitem{shot_Chen_Natelson_arxiv_2023}
	L. Chen, D. T. Lowder, E. Bakali, A. M. Andrews, W. Schrenk, M. Waas, R. Svagera, G. Eguchi, L. Prochaska, Y. Wang, C. Setty, S. Sur, Q. Si, S. Paschen, D. Natelson, 
	Shot noise indicates the lack of quasiparticles in a strange metal, 
	Science \textbf{382}, 907 (2023).
\bibitem{MFL_Wu_Liao_Foster_PRB_22}
	T. C. Wu, Y. Liao, and M. S. Foster,
	Quantum Interference of Hydrodynamic Modes in a Dirty Marginal Fermi Liquid,
	Phys. Rev. B \textbf{106}, 155108 (2022).
\bibitem{disO_SC_MFC_Wu_Lee_Foster_arxiv_2023}
	T. C. Wu, P. A. Lee, M. S. Foster,
	Enhancement of Superconductivity in a Dirty Marginal Fermi Liquid,
	Phys. Rev. B \textbf{108}, 214506 (2023).
\bibitem{Sonner2012}
	J. Sonner and A. G. Green,
	Hawking Radiation and Nonequilibrium Critical Current Noise,
	Phys. Rev. Lett. {\bf 109}, 091601 (2012).
\bibitem{supplemental_material}
	See the Supplemental Material for
	the Keldysh computation of the diagrams,
	the derivation and solution to the kinetic equation,
	the derivation of Eq.~(\ref{eq:shot_eqm_boson_T=0}),
	and the comparison of experimental data from Ref.~\cite{shot_Chen_Natelson_arxiv_2023} to Eq.~(\ref{TcrossoverQuarter}),
	and which includes Refs.~\cite{Wu_Pal_Foster_WSC_PRB_2021}--\cite{NLsM5_Finkelshtein_83}.
\bibitem{SYK_Patel_PRB_21}
	I. Esterlis, H. Guo, A. A. Patel, and S. Sachdev, 
	Large-$N$ theory of critical Fermi surfaces, 
	Phys. Rev. B \textbf{103}, 235129 (2021).
\bibitem{SYK_Guo_Patel_linearT_PRB_2022}
	H. Guo, A. A. Patel, I. Esterlis, S. Sachdev,
	Large $N$ theory of critical Fermi surfaces II: conductivity,
	Phys. Rev. B \textbf{106}, 115151 (2022).
\bibitem{SYK_Patel_linearT_arxiv_2022}
	A. A. Patel, H. Guo, I. Esterlis, S. Sachdev,
	Universal theory of strange metals from spatially random interactions,
	Science {\bf 381}, 790 (2023).
\bibitem{disO_SC_MFC_Sri_Burmistrov_PRB_2023}
	P. A. Nosov, I. S. Burmistrov, and S. Raghu, 
	Interplay of superconductivity and localization near a 2D ferromagnetic quantum critical point, 
	Phys. Rev. B {\bf 107}, 144508 (2023).
\bibitem{Galitski05}
	V. M. Galitski,
	Metallic phase in a two-dimensional disordered Fermi system with singular interactions,
	Phys. Rev. B {\bf 72}, 214201 (2005).
\bibitem{thermal_mass_Torroba}
	J. A. Damia, M. Sol\'is, and G. Torroba,
	How non-Fermi liquids cure their infrared divergences, 	
	Phys. Rev. B \textbf{102}, 045147 (2020).
\bibitem{MFL_Varma_CuO_PRL_89}
	C. M. Varma, P. B. Littlewood, S. S.-Rink, E. Abrahams, and A. E. Ruckenstein, 
	Phenomenology of the normal state of Cu-O high-temperature superconductors, 	
	Phys. Rev. Lett. \textbf{63}, 1996 (1989).
\bibitem{NFL_SSLee_Review_18}
	S.-S. Lee, 
	Recent Developments in Non-Fermi Liquid Theory, 
	Ann. Rev. of Cond. Matt. Phys., \textbf{9}, 227 - 244 (2018).
\bibitem{NFL_Subir_book_CUP_2011}
	S. Sachdev,
	\textit{Quantum Phase Transitions}, 2nd Ed., 
	Cambridge University Press (2011). 
\bibitem{NFL_SU_N_Raghu_PRL_19}
	J. A. Damia, S. Kachru, S. Raghu, and G. Torroba, 
	Two-Dimensional Non-Fermi-Liquid Metals: A Solvable Large-$N$ Limit,
	Phys. Rev. Lett. \textbf{123}, 096402 (2019).
\bibitem{Keldysh_conv4_Kamenev_AdvPhy_09}
	A. Kamenev and A. Levchenko, 
	Keldysh technique and non-linear $\sigma$-model: basic principles and applications, 
	Adv. Phys. \textbf{58}, 197 (2009).
\bibitem{NFL_SU_N_disO_Raghu_PRL_20}
	P. A. Nosov, I. S. Burmistrov, and S. Raghu, 
	Interaction-Induced Metallicity in a Two-Dimensional Disordered Non-Fermi Liquid, 
	Phys. Rev. Lett. \textbf{125}, 256604 (2020).
\bibitem{10-fold_Altland_Zirnbauer_PRB_1997}
	A. Altland and M. R. Zirnbauer, 
	Nonstandard symmetry classes in mesoscopic normal-superconducting hybrid structures,	
	Phys. Rev. B \textbf{55}, 1142 (1997).
\bibitem{10-fold_Ludwig_NewJPhy_2010}
	S. Ryu, A. P. Schnyder, A. Furusaki, and A. W. W. Ludwig,
	Topological insulators and superconductors: tenfold way and dimensional hierarchy, 
	New J. Phys. \textbf{12}, 065010 (2010).
\bibitem{SYK_review_Sachdev_RMP_2022}
	D. Chowdhury, A. Georges, O. Parcollet, and S. Sachdev, 
	Sachdev-Ye-Kitaev Models and Beyond: A Window into Non-Fermi Liquids, 
	Rev. Mod. Phys. \textbf{94}, 035004 (2022). 
\bibitem{NLsM2_Foster_Liao_Ann_17}
	Y. Liao, A. Levchenko, and M. S. Foster,
	Response theory of the ergodic many-body delocalized phase: Keldysh Finkel'stein sigma models and the 10-fold way, 
	Ann. Phys. \textbf{386}, 97 (2017).
\bibitem{shot_AA_Gefen_PRB_2001}
	D. B. Gutman and Y. Gefen, 
	Shot noise in disordered junctions: Interaction corrections, 
	Phys. Rev. B \textbf{64}, 205317 (2001).
\bibitem{Wu_Pal_Foster_WSC_PRB_2021}
	T. C. Wu, H. K. Pal, and M. S. Foster,
	Topological anomalous skin effect in Weyl superconductors,
	Phys. Rev. B \textbf{103}, 104517 (2021). 
\bibitem{Chubukov_MT_PRB_2012}
	D. L. Maslov and A. V. Chubukov,
	First-Matsubara-frequency rule in a Fermi liquid. II. Optical conductivity and comparison to experiment,
	Phys. Rev. B \textbf{86}, 155137 (2012).
\bibitem{NLsM5_Finkelshtein_83}
	A. M. Finkel'stein,
	Influence of Coulomb interaction on the properties of disordered metals, 
	Zh. Eksp. Teor. Fiz. \textbf{84}, 168 (1983)
	[Sov. Phys. JETP {\bf 57}, 97 (1983)].
\bibitem{shot_Sachdev_arxiv_2023}
	A. Nikolaenko, S. Sachdev, A. A. Patel,
	Theory of shot noise in strange metals,
	Phys. Rev. Research \textbf{5}, 043143 (2023).
\end{thebibliography}
\end{document}


\title{
Suppression of Shot Noise in a Dirty Marginal Fermi Liquid: 
\\
Supplemental Material
}

\author{Tsz Chun Wu}
\affiliation{Department of Physics and Astronomy, Rice University, Houston, Texas 77005, USA}
\author{Matthew S. Foster}
\affiliation{Department of Physics and Astronomy, Rice University, Houston, Texas 77005, USA}
\affiliation{Rice Center for Quantum Materials, Rice University, Houston, Texas 77005, USA}

\date{\today}

\maketitle
\tableofcontents


\section{Evaluation of the bubble and Maki-Thompson diagrams for shot noise}

In this section, we derive the shot noise contributed by the bubble and Maki-Thompson (MT) diagrams \blue{(Fig.~3)} using the Keldysh path integral formalism. 
Our conventions closely follow our previous works in Refs.~\cite{MFL_Wu_Liao_Foster_PRB_22,disO_SC_MFC_Wu_Lee_Foster_arxiv_2023,Wu_Pal_Foster_WSC_PRB_2021}.
The partition function $Z[A]$ in this work is essentially \blue{Eq.~(26)} in Ref.~\cite{MFL_Wu_Liao_Foster_PRB_22}. 
However, instead of introducing source field $V$ that couples with the fermionic density operator, we introduce a quantum component of the vector potential source field $A^{\sfq}$ that couples with the classical component of the electric current operator \cite{Wu_Pal_Foster_WSC_PRB_2021,NLsM6_Kamenev_CUP_11}. 

The shot noise can be evaluated by computing the Keldysh component of the current-current correlator \cite{NLsM6_Kamenev_CUP_11}:
\begin{equation}
	S_{\mathsf{shot}}
	=
	-
	\frac{e^2}{2}
	\frac{\delta^2 \ln Z[A]}{\delta A^{\sfq}_{\Omega} \, \delta A^{\sfq}_{-\Omega}}
	\bigg|_{A^{\sfq} = 0, \Omega \rightarrow 0}
	=
	S_{\sfB}
	+
	S_{\sfMT}
	+
	\text{quantum corrections},
\end{equation} 
which generates the bubble and MT diagrams at the semiclassical level. 
We leave the quantum interference corrections arising from weak localization \cite{NLsM6_Kamenev_CUP_11,NLsM2_Foster_Liao_Ann_17} and Altshuler-Aronov contributions \cite{shot_AA_Gefen_PRB_2001,MFL_Wu_Liao_Foster_PRB_22} for further studies.

\subsection{Bubble diagram}

The bubble diagram is given by
\begin{equation}
	S_{\sfB}
	=
	\lim\limits_{\Omega \rightarrow 0}
	\frac{Ne^2 \, v_F^2}{2d}
	\int_0^L \frac{dx}{L}
	\int\limits_{\omega,\textbf{k}}
	\Tr
	\left[
		\hG_{\omega + \Omega}(\textbf{k}) 
		\hgammaq_{\omega + \Omega,\omega} \hG_{\omega}(\textbf{k}) 
		\hgammaq_{\omega,\omega + \Omega} 
	\right],
\end{equation}
where $\int_{\omega} = \int_{-\infty}^{\infty} \frac{d\omega}{2\pi}$, $\int_{\textbf{k}} = \int \frac{d^2 \textbf{k}}{(2\pi)^2}$, the dimension $d = 2$, and $v_F$ is the Fermi velocity, and the Green's function is
\begin{equation}
	\hG_{\omega,\textbf{k}}
	=
	\begin{bmatrix}
		G^R_{\omega,\textbf{k}} & 0
		\\
		0 & G_{\omega,\textbf{k}}^A
	\end{bmatrix}_{\tau},
\end{equation}
with $G^{R/A}_{\omega,\textbf{k}}$ being the retarded/advanced Green's function. 
The subscript $\tau$ indicates that the matrix is expressed in the Keldysh $\tau$ space. 
At the saddle point level \cite{MFL_Wu_Liao_Foster_PRB_22}, 
\begin{equation}
	G^{R}_{\omega,\textbf{k}}
	=
	\frac{1}{
		\omega + i\gammael - \tilde{\veps}_{\textbf{k}} - \Sigma^R_{\omega}
	}
	,
\end{equation}
where $\tilde{\veps}_{\textbf{k}} = \textbf{k}^2/2m - \mu$, $m$ is the electron mass, $\mu$ is the chemical potential, and $\Sigma^R_{\omega}$ is the retarded MFL self-energy \cite{MFL_Wu_Liao_Foster_PRB_22}. 
The trace runs over the Keldysh space $\tau$, $\hgammacl_{\omega,\omega'} = \hM_{F,\omega} \hM_{F,\omega'}$, $\hgammaq_{\omega,\omega'} = \hM_{F,\omega} \htau^1 \hM_{F,\omega'}$, and the thermal matrix $\hM_{F,\omega} = 
\begin{bmatrix}
1 & F_{\omega} \\ 0 & -1
\end{bmatrix}_{\tau}$.
The external frequency $\Omega$ is sent to zero for shot noise. 

Expanding the trace and ignoring purely retarded/advanced contributions, we have 
\begin{equation}
	S_{\sfB}
	=
	\sigma_{\mathsf{D}}
	\int_0^L \frac{dx}{L}
	\int
	d\omega
	\frac{1}{1 -   \im \Sigma^R_{\omega}(x)/\gammael}
	\left[
		1 - F_{\omega}^2(x)
	\right]
	,
\end{equation}
which is \blue{Eq.~(8)} in the main text. 
We have converted $\int_{\textbf{k}} = \nu_0 \int d\tilde{\veps}_{\textbf{k}}$, where $\nu_0$ is the density of states on the Fermi surface.

\subsection{Maki-Thompson diagram}

Consider the Maki-Thompson diagram contribution to shot noise
\begin{equation}
	S_{\sfMT}
	=
	\frac{i Ne^2}{2d}
	\int_0^L \frac{dx}{L}
	I_{\sfMT}(x)
\end{equation}
where $d = 2$ and 
\begin{equation}
\begin{aligned}
	&I_{\mathsf{MT}}(x)
	\\
	&=
	\lim_{\Omega \rightarrow 0}
	g^2 
	\int\limits_{\textbf{k},\textbf{q},\omega,\nu}
	\textbf{v}_{F,\textbf{k}} \cdot \textbf{v}_{F,\textbf{k}+\textbf{q}}
	\,
	D^{ss'}_{-\nu,-\textbf{q}}
	\Tr
	\begin{bmatrix}
		\hgammacl_{\omega + \Omega,\omega}
		\hG_{\omega,\textbf{k}} 
		\hgamma^s_{\omega,\omega + \nu} 
		\hG_{\omega + \nu,\textbf{k} + \textbf{q}} 
		\hgammaq_{\omega + \nu,\omega + \nu + \Omega} 
		\hG_{\omega + \nu + \Omega,\textbf{k} + \textbf{q}}
		\hgamma^{s'}_{\omega + \nu + \Omega,\omega + \Omega}
		\hG_{\omega + \Omega,\textbf{k}}
	\end{bmatrix}
	,
\end{aligned}
\end{equation}
where  $s,s' \in \left\lbrace \sfcl,\sfq \right\rbrace$,  
\begin{equation}
	\hD_{\nu,\textbf{q}}
	=
	\begin{bmatrix}
		D^K_{\nu,\textbf{q}} & D^R_{\nu,\textbf{q}}
		\\
		D^A_{\nu,\textbf{q}} & 0
	\end{bmatrix}_{\tau}, 
	\qquad
	D^K_{\nu,\textbf{q}}
	= 
	(D^R_{\nu,\textbf{q}} - D^A_{\nu,\textbf{q}}) F_{B,\nu},
	\qquad
	D^R_{\nu,\textbf{q}}
	=
	-
	\frac{1}{2(q^2 - i\alpha \, \nu + \mb^2)},
	\qquad
	\mb^2 = \alpha_m T. 
\end{equation}
Expanding the $\Tr$ by brute force and retaining terms that do not vanish under causality, we have 
\begin{equation}
\begin{aligned}
	I_{\mathsf{MT}}(x)
	&=
	2
	g^2
	\int\limits_{\textbf{k},\textbf{q},\omega,\nu}
	\textbf{v}_{F,\textbf{k}} \cdot \textbf{v}_{F,\textbf{k}+\textbf{q}}
	(D^R_{-\nu,-\textbf{q}} - D^A_{-\nu,-\textbf{q}})
	G^R_{\omega,\textbf{k}}
	G^A_{\omega,\textbf{k}}
	G^R_{\omega + \nu,\textbf{k} + \textbf{q}}
	G^A_{\omega + \nu,\textbf{k} + \textbf{q}}
	\,
	\calF (\omega,\nu)
	\\
	&=
	2i
	g^2
	\int\limits_{\textbf{k},\textbf{q},\omega,\nu}
	\textbf{v}_{F,\textbf{k}} \cdot \textbf{v}_{F,\textbf{k}+\textbf{q}}
	[
		2\im D^R_{-\nu,-\textbf{q}} 
	]
	G^R_{\omega,\textbf{k}}
	G^A_{\omega,\textbf{k}}
	G^R_{\omega + \nu,\textbf{k} + \textbf{q}}
	G^A_{\omega + \nu,\textbf{k} + \textbf{q}}
	\,
	\calF (\omega,\nu)
,
\end{aligned}
\end{equation}
where the thermal factor
\begin{equation}
\label{eq:MT_thermal_F}
	\calF (\omega,\nu)
	=
	F_{B,-\nu}
	\left(
		F_{\omega} F_{\omega + \nu} 
		-
		 F_{\omega + \nu }^2
		+
		1
		-
		 F_{\omega}^2
	\right)
	+
	F_{\omega} F_{\omega + \nu}
	\left(
		F_{\omega + \nu}- F_{\omega}
	\right)
.
\end{equation}
In \blue{Eq.~(11)} of the main text, we ignore finite angle scattering contributions by approximating 
$\textbf{v}_{F,\textbf{k}} \cdot \textbf{v}_{F,\textbf{k}+\textbf{q}} \simeq v_F^2$.


\section{Fluctuation-dissipation theorem and the Johnson–Nyquist noise}
\label{sec:FDT}

In this section, we verify our expression for the shot noise reduces to the Johnson–Nyquist noise at equilibrium, as required by the fluctuation-dissipation theorem. 
Recall that at equilibrium, the fermionic and bosonic distribution functions become
\begin{equation}
	F_{\omega} = \tanh\left(\frac{\omega}{2T}\right)
	,
	\qquad
	F_{B,\Omega} = \coth\left(\frac{\Omega}{2T}\right).
\end{equation}

For the bubble diagram, 
\begin{equation}
\label{eq:shot_B_eqm}
\begin{aligned}
	S_{\sfB}
	&=
	\sigma_{\mathsf{D}}
	\int
	d\omega
	\frac{1}{1 -   \im \Sigma^R_{\omega}(x)/\gammael}
	\sech^2\left(\frac{\omega}{2T}\right)
	\\
	&=
	\lim_{\Omega \rightarrow 0}
	\sigma_{\mathsf{D}}
	\int
	d\omega
	\frac{1}{1 -   \im \Sigma^R_{\omega}(x)/\gammael}
	\frac{1}{2T}
	\sech^2\left(\frac{\omega}{2T}\right)
	\,
	\Omega \, \coth\left( \frac{\Omega}{2T}\right). 
\end{aligned}
\end{equation}

For the MT diagram, the thermal factor at equilibrium becomes
\begin{equation}
\begin{aligned}
	\calF (\omega,\nu)
	&=
	-F_{B,\nu}
	\left(
		F_{\omega} F_{\omega + \nu} 
		-
		F_{\omega + \nu }^2
		+
		1
		-
		 F_{\omega}^2
	\right)
	+
	F_{\omega } F_{\omega + \nu}
	\left(
		F_{\omega + \nu}- F_{\omega}
	\right)
	\\
	&=
	-F_{B,\nu} (1 - F_{\omega}^2)
	+
	F_{B,\nu}
	 F_{\omega + \nu} 
	\left(
		 F_{\omega + \nu} 
		-
		 F_{\omega}
	\right)
	+
	F_{\omega } F_{\omega + \nu}
	\left(
		 F_{\omega + \nu}- F_{\omega}
	\right)
	\\
	&=
	-F_{B,\nu} \sech^2\left(\frac{\omega}{2T}\right)
	+
	F_{B,\nu}
	 F_{\omega + \nu} 
	\left(
		 F_{\omega + \nu} 
		-
		 F_{\omega}
	\right)
	+
	[
		1 - F_{B,\nu}(F_{\omega + \nu} - F_{\omega})
	]
	\left(
		 F_{\omega + \nu}- F_{\omega}
	\right)
	\\
	&=
	-F_{B,\nu} \sech^2\left(\frac{\omega}{2T}\right)
	+
	(
		1
		+
		F_{B,\nu}F_{\omega}
	)
	\left(
		 F_{\omega + \nu}- F_{\omega}
	\right)
	\\
	&=
	-F_{B,\nu} \sech^2\left(\frac{\omega}{2T}\right)
	+
	( F_{\omega + \nu}- F_{\omega})
	+
	F_{\omega}
	(
		1 - F_{\omega}F_{\omega + \nu}
	)
	\\
	&=
	(F_{\omega + \nu} -F_{B,\nu}) \sech^2\left(\frac{\omega}{2T}\right)
	\\
	&=
	\lim_{\Omega \rightarrow 0}
	(F_{\omega + \nu} -F_{B,\nu}) 
	\frac{1}{2T}
	\sech^2\left(\frac{\omega}{2T}\right)
	\,
	\Omega \, \coth\left(\frac{\Omega}{2T}\right),
\end{aligned}
\end{equation}
where we have used the identity
$F_{\omega} F_{\omega + \nu}
=
1 - F_{B,\nu}(F_{\omega + \nu} - F_{\omega})$. 
Thus, 
\begin{equation}
	S_{\sfMT}
	=
	\lim_{\Omega \rightarrow 0}
	\frac{e^2}{d}
	g^2
	\int\limits_{\textbf{k},\textbf{q},\omega,\nu}
	\textbf{v}_{F,\textbf{k}} \cdot \textbf{v}_{F,\textbf{k}+\textbf{q}}
	[2\im D^R_{\nu,\textbf{q}} ]
	G^R_{\omega,\textbf{k}}
	G^A_{\omega,\textbf{k}}
	G^R_{\omega + \nu,\textbf{k} + \textbf{q}}
	G^A_{\omega + \nu,\textbf{k} + \textbf{q}}
	\,
	(F_{\omega + \nu} -F_{B,\nu}) 
	\frac{1}{2T}
	\sech^2\left(\frac{\omega}{2T}\right)
	\,
	\Omega \, \coth\left(\frac{\Omega}{2T}\right),
\end{equation}
which is reminiscent of the imaginary part of the retarded MT current-current correlation function \cite{Chubukov_MT_PRB_2012} multiplied by $\coth(\Omega/2T)$, as expected by the fluctuation-dissipation theorem. 

To proceed, we approximate 
\begin{equation}
\label{eq:shot_MT_eqm}
\begin{aligned}
	S_{\sfMT}
	&\simeq
	\lim_{\Omega \rightarrow 0}
	\frac{Ne^2}{d}
	g^2
	\frac{v_F^2}{\gammael^2}
	\int\limits_{\textbf{k},\textbf{q},\omega,\nu}
	2\im D^R_{\nu,\textbf{q}} 
	G^R_{\omega,\textbf{k}}
	G^A_{\omega,\textbf{k}}
	\,
	(F_{\omega + \nu} -F_{B,\nu}) 
	\frac{1}{2T}
	\sech^2\left(\frac{\omega}{2T}\right)
	\,
	\Omega \, \coth\left(\frac{\Omega}{2T}\right)
\\
	&\simeq
	\lim_{\Omega \rightarrow 0}
	\frac{Ne^2}{d}
	g^2
	\frac{v_F^2}{\gammael^2}
	\int\limits_{\omega,\nu}
	\frac{-1}{4\pi}
	\tan^{-1}\left(\frac{\alpha \nu}{\mb^2}\right)
	\frac{\pi \nu_0}{
		\gammael - \im \Sigma^R_{\omega}
	}
	\,
	(F_{\omega + \nu} -F_{B,\nu}) 
	\frac{1}{2T}
	\sech^2\left(\frac{\omega}{2T}\right)
	\,
	\Omega \, \coth\left(\frac{\Omega}{2T}\right)
\\
	&=
	-
	\lim_{\Omega \rightarrow 0}
	\sigma_{\mathsf{D}}
	\frac{\gbar^2}{2 }
	\int d\omega
	\frac{1 }{
		\gammael - \im \Sigma^R_{\omega}
	}
	\int
	 d\nu
	\tan^{-1}\left(\frac{\alpha \nu}{\mb^2}\right)
	\,
	(F_{\omega + \nu} -F_{B,\nu}) 
	\frac{1}{2T}
	\sech^2\left(\frac{\omega}{2T}\right)
	\,
	\Omega \, \coth\left(\frac{\Omega}{2T}\right)
\\
	&=
	-
	\lim_{\Omega \rightarrow 0}
	\sigma_{\mathsf{D}}
	\int d\omega
	\frac{\im \Sigma^R_{\omega}/\gammael}{
		1 - \im \Sigma^R_{\omega} /\gammael
	}
	\frac{1}{2T}
	\sech^2\left(\frac{\omega}{2T}\right)
	\,
	\Omega \, \coth\left(\frac{\Omega}{2T}\right),
\end{aligned}
\end{equation}
where we have used the definition of the self-energy \blue{[Eq.~(9)]} in the main text.

As a result, by combining Eqs.~(\ref{eq:shot_B_eqm}) and (\ref{eq:shot_MT_eqm}), the shot noise at equilibrium is
\begin{equation}
\begin{aligned}
	\shot 
	&=
	S_{\sfB} + S_{\sfMT}
	\\
	&=
	\lim_{\Omega \rightarrow 0}
	\sigma_{\mathsf{D}}
	\int d\omega
	\frac{1}{2T}
	\sech^2\left(\frac{\omega}{2T}\right)
	\,
	\Omega \, \coth\left(\frac{\Omega}{2T}\right)
	\\
	&=
	\lim_{\Omega \rightarrow 0}
	\sigma_{\mathsf{D}}
	2
	\Omega \, \coth\left(\frac{\Omega}{2T}\right)
	\\
	&=
	4T \sigma_{\mathsf{D}},
\end{aligned}
\end{equation}
which is the Johnson–Nyquist noise. 
Notice that the MFL self-energy correction is canceled, as in the electric conductivity calculations \cite{MFL_Wu_Liao_Foster_PRB_22,SYK_Guo_Patel_linearT_PRB_2022,SYK_Patel_linearT_arxiv_2022}. 
However, we emphasize that this is only the equilibrium result. 
In the following, we will see that the interaction correction does not drop out in a non-equilibrium situation in general.


\section{Derivation of the kinetic equation}

The field theoretical description of a system of disordered interacting electrons in the diffusive regime is given by the Finkel’stein nonlinear sigma model (FNLsM) \cite{NLsM2_Foster_Liao_Ann_17,NLsM5_Finkelshtein_83,NLsM6_Kamenev_CUP_11,Keldysh_conv4_Kamenev_AdvPhy_09}. 
We follow the MFL version of the FNLsM presented in our recent work \cite{MFL_Wu_Liao_Foster_PRB_22} within the Keldysh formalism to derive the kinetic equation governing the distribution function.   
We restrict our attention in this work to the particle-hole channel (unitary class \cite{10-fold_Altland_Zirnbauer_PRB_1997,10-fold_Ludwig_NewJPhy_2010}) and ignore the Cooperon contributions. 

Consider the $\hQ$-matrix action in the MFL-FNLsM \cite{MFL_Wu_Liao_Foster_PRB_22}
\begin{equation}
\begin{aligned}
	S_{Q}
	&=
	\frac{\pi \nu_0}{4}
	\int\limits_{\textbf{x}}
	\left[
		D
		\Tr (\nabla \hQ \nabla \hQ)
		+
		4i  \, \Tr [\hQ \,(\hat{\omega} - \hSig_{\homega})]
	\right]
	,
\end{aligned}
\end{equation}
where 
\begin{equation}
	\hQ = \hU^{-1} \hQ_{\sfsp} \hU,
	\qquad
	\hU = e^{\hW/2}
	,
	\qquad 
	\hQ_{\sfsp;t_1,t_2}(\textbf{x})
	=
	\begin{bmatrix}
		1 & 2F_{t_1,t_2}(\textbf{x})
		\\
		0 & -1
	\end{bmatrix}_{\tau}
	=
	\hat{M}_F(\textbf{x})
	\hqsp
	\hat{M}_F(\textbf{x})
	,
	\qquad
	\hqsp = \htau^3
	,
\end{equation}
$\hW$ is the generator of rotation, and 
\begin{equation}
\hSig_{\homega}
=
\begin{bmatrix}
\Sigma^R_{\omega} & \Sigma^K_{\omega}
\\
0 & \Sigma^A_{\omega}
\end{bmatrix}_{\tau}
.
\end{equation}
Here, $\Sigma^{R/A/K}_{\omega}$ denotes the retarded/advanced/Keldysh component of the MFL self-energy. 
We adopt the $\hQ$-matrix convention in Ref.~\cite{NLsM6_Kamenev_CUP_11}
which incorporates the fermion distribution function into the matrix field,
instead of using $\hq$ as in our previous works \cite{MFL_Wu_Liao_Foster_PRB_22,disO_SC_MFC_Wu_Lee_Foster_arxiv_2023}. 
The diffusion constant $D$ is assumed to be space-time independent. 
We focus on the implication of $\hSig_{\homega}$ for the kinetic equation and leave quantum corrections arising from the interaction terms $S_{\sfintI}$ and $S_{\sfintII}$ \cite{MFL_Wu_Liao_Foster_PRB_22} for further studies.  

Expanding to linear order of $\hW$, we have
\begin{equation}
\hQ = \hQ_{\sfsp} - \frac{1}{2}[\hW, \hQ_{\sfsp}] + {\cal O}(\hW^2).
\end{equation}
Using the cyclic property of $\Tr$ and integration by parts, the gradient term can be written as
\begin{align}
	D
	\Tr (\nabla \hQ \nabla \hQ)
	=
	-\Tr
	\left[
		D 
		\nabla \hQ_{\sfsp}
		\nabla (\hW \hQ_{\sfsp}  - \hQ_{\sfsp} \hW )
	\right]
	=
	2
	\,
	\Tr
	\left[
		\hW \nabla ( D\hQ_{\sfsp} \nabla \hQ_{\sfsp})
	\right].
\end{align}
The frequency term is
\begin{align}
	4i
	\,
	\Tr
	\left[
		\hat{\omega} \hQ
	\right]
	=
	-2
	\,
	\int
	\Tr
	\left[
		\hW_{t_1,t_2}(\partial_{t_1}  + \partial_{t_2})\hQ_{\sfsp;t_2,t_1}
	\right]
	.
\end{align}
The above derivation closely follows Ref. \cite{NLsM6_Kamenev_CUP_11}. 
Modification to the kinetic equation arises from the self-energy term $\hSig_{\homega}$:
\begin{align}
	-4i 
	\,
	\Tr[ \hSig \;\hQ]
	=
	2i
	\,
	\Tr
	\left[
		\hW(
		\hQ_{\sfsp} \hSig
		-
		\hSig \hQ_{\sfsp}
		)
	\right]
	.
\end{align}
By varying with respect to $\hW$ and focusing on the $(1,2)$ component of the matrix equation, we obtain the kinetic equation
\begin{equation}
	-D \, \nabla^2 F_{\omega}(t,\textbf{x})
	+
	\partial_t F_{\omega}(t,\textbf{x})
	=
	i 
	\left[
		\Sigma^K
		+
		F \Sigma^A
		-
		\Sigma^R F
	\right]
	=
	\St_{\sfeb}[F]
	,
\end{equation}
where we have employed Wigner transformation
\begin{equation}
	F_{t_1,t_2}(\textbf{x})
	=
	\int\limits_{\veps} F_{\omega}(t,\textbf{x})\; e^{-i \omega\, \delta t}
	,
	\qquad
	t = \frac{t_1 + t_2}{2}
	,
	\qquad
	\delta t = t_1 - t_2
	.
\end{equation}
Using \cite{MFL_Wu_Liao_Foster_PRB_22,NLsM6_Kamenev_CUP_11}
\begin{eqnarray}
	\Sigma^{R/A}_{\omega,\textbf{p}}(t,\textbf{x})
	&=&
	i \, g^2
	\int\limits_{\Omega,\textbf{q}}
	\left[
		G^{R/A}_{\omega + \Omega,\textbf{p} + \textbf{q}}(t,\textbf{x})
		D^K_{\Omega,\textbf{q}}(t,\textbf{x})
		+
		G^{K}_{\omega + \Omega,\textbf{p} + \textbf{q}}(t,\textbf{x})
		D^{A/R}_{\Omega,\textbf{q}}(t,\textbf{x})
	\right]
	\\
	\Sigma^{K}_{\omega,\textbf{p}}(t,\textbf{x})
	&=&
	i \, g^2
	\int\limits_{\Omega,\textbf{q}}
	\left[
		G^{K}_{\omega + \Omega,\textbf{p} + \textbf{q}}(t,\textbf{x})
		D^K_{\Omega,\textbf{q}}(t,\textbf{x})
		-
		[G^{R} - G^A]_{\omega + \Omega,\textbf{p} + \textbf{q}}(t,\textbf{x})
		[D^{R}-D^A]_{\Omega,\textbf{q}}(t,\textbf{x})
	\right]
	,
\end{eqnarray}
where $G^K = (G^R - G^A) F$ and $D^K = (D^R - D^A) F_B$, 
the collision integral can be written as
\begin{equation}
	\St_{\sfeb}[F]
	=
	4g^2
	\int\limits_{\Omega,\textbf{q}}
	\im G^{R}_{\omega + \Omega,\textbf{p} + \textbf{q}}(t,\textbf{x})
	\;
	\im D^{R}_{\Omega,\textbf{q}}(t,\textbf{x})
	\begin{Bmatrix}
		F_{B,\Omega}(t,\textbf{x})
		[
			F_{\omega + \Omega}(t,\textbf{x})
			-
			F_{\omega}(t,\textbf{x})
		]
		-
		(
			1 
			-
			F_{\omega + \Omega}(t,\textbf{x})
			F_{\omega}(t,\textbf{x})
		)
	\end{Bmatrix}
	,
\end{equation}
which is \blue{Eq.~(14)} in the main text.
At equilibrium, $F_{\omega} = \tanh(\omega/2T)$ and $F_{B,\Omega} = \coth(\Omega/2T)$, the thermal factor inside $\left\lbrace ... \right\rbrace$ vanishes due to the identity
\begin{equation}
	1 - \tanh x \tanh y
	=
	\coth (x - y)
	(\tanh x  - \tanh y). 
\end{equation}

\subsection{Equilibrium boson regime}

We now assume that the bosons are at equilibrium, i.e. $F_{B,\Omega} = \coth\left( \frac{\Omega}{2T} \right)$ and focus on time-independent distribution functions. 
In the regime $l_{\sfel} \ll l_{\sfeb} \ll L$, we consider a fermionic distribution function of the form
\begin{equation}
	F_{\omega}(x)
	=
	\tanh 
	\left[
	\frac{
		\omega - e \, V \, \xbar
	}{
		2T_{\mathsf{e}}(\xbar)
	}\right],
\end{equation} 
where $\xbar = x/L$. 
We approximate the collision integral by
\begin{equation}
\begin{aligned}
	\St_{\sfeb}[F]
	&\simeq
	2g^2
	\int\limits_{\Omega,\textbf{q}}
	\frac{1}{\gammael}
	\frac{
		\alpha \, \Omega
	}{
		(q^2 + \mb^2)^2 + (\alpha \, \Omega)^2
	}
	\begin{Bmatrix}
		F_{B,\Omega}(x)
		[
			F_{\omega + \Omega}(x)
			-
			F_{\omega}(x)
		]
		-
		[
			1 
			-
			F_{\omega + \Omega}(x)
			F_{\omega}(x)
		]
	\end{Bmatrix}
	\\
	&=
	2g^2
	\int\limits_{\Omega}
	\int_0^{\infty}
	\frac{dx}{4\pi}
	\frac{1}{\gammael}
	\frac{\alpha \, \Omega}{
		(x + \mb^2)^2 + (\alpha \, \Omega)^2
	}
	\begin{Bmatrix}
		F_{B,\Omega}(t,\textbf{x})
		[
			F_{\omega + \Omega}(x)
			-
			F_{\omega}(x)
		]
		-
		[
			1 
			-
			F_{\omega + \Omega}(x)
			F_{\omega}(x)
		]	
	\end{Bmatrix}
	\\
	&=
	\frac{2g^2}{(4\pi\gammael)(2\pi)}
	\int d\Omega
	\,
	\tan^{-1}
	\left(\frac{\alpha \, \Omega}{\mb^2}\right)
	\begin{Bmatrix}
		F_{B,\Omega}(x)
		[
			F_{\omega + \Omega}(x)
			-
			F_{\omega}(x)
		]
		-
		[
			1 
			-
			F_{\omega + \Omega}(x)
			F_{\omega}(x)
		]
	\end{Bmatrix}
	\\
	&=
	\gbar^2
	\int d\Omega
	\,
	\tan^{-1}
	\left(\frac{\alpha \, \Omega}{\mb^2}\right)
	\begin{Bmatrix}
		F_{B,\Omega}(x)
		[
			F_{\omega + \Omega}(x)
			-
			F_{\omega}(x)
		]
		-
		[
			1 
			-
			F_{\omega + \Omega}(x)
			F_{\omega}(x)
		]
	\end{Bmatrix}
	,
\end{aligned}
\end{equation}
where $\gbar^2 = g^2/(4\pi^2\gammael)$.

The kinetic equation is
\begin{equation}
\label{eq:Te_ode}
	\frac{\pi^2}{6}
	\nabla_{\bar{x}}^2 
	\Te^2(\bar{x})
	=
	-
	(e V)^2
	+
	\frac{1}{2E_{\mathsf{Th}}}
	\int_{-\infty}^{\infty} 
	d\omega
	\,
	\omega
	\,
	\St_{\sfeb}[F],
\end{equation}
where the collision integral term is
\begin{equation}
\begin{aligned}[b]
	\int_{-\infty}^{\infty} d\omega
	\,
	\omega
	\,
	\St_{\sfeb}[F]
&=
	\bar{g}^2
	\int_{-\infty}^{\infty} d\omega
	\int_{-\infty}^{\infty} d\Omega
	\,
	\omega
	\tan^{-1}
	\left(\frac{\alpha \, \Omega}{\mb^2}\right)
	\begin{Bmatrix}
	F_{B,\Omega}(x)
	[
		F_{\omega + \Omega}(x)
		-
		F_{\omega}(x)
	]
	-
	[
		1 
		-
		F_{\omega + \Omega}(x)
		F_{\omega}(x)
	]	\end{Bmatrix}
\\
&=
	\bar{g}^2
	\int_{-\infty}^{\infty} d\omega
	\int_{-\infty}^{\infty} d\Omega
	\,
	\omega
	\tan^{-1}
	\left(\frac{\alpha \, \Omega}{\alpha_m T}\right)
\begin{aligned}[t]
&
	\left\{
		\coth\left[\frac{\Omega}{2T}\right]
		-
		\coth\left[\frac{\Omega}{2\Te(\bar{x})}\right]
	\right\}
\\
&
\times
	\left\{
		\tanh\left[\frac{\omega + \Omega - V \bar{x}}{2\Te(\bar{x})}\right]
		-
		\tanh\left[\frac{\omega  - V \bar{x}}{2\Te(\bar{x})}\right]
	\right\}
\end{aligned}
\\
&=
	-
	2\bar{g}^2
	\int_{0}^{\infty} d\Omega
	\,
	\tan^{-1}
	\left(\frac{\alpha \, \Omega}{\alpha_m T}\right)
	\left\{
	\coth\left[ \frac{\Omega}{2T} \right]
	-
	\coth\left[\frac{\Omega}{2\Te(\bar{x})} \right]
	\right\}
	\Omega^2,
\end{aligned}
\end{equation}
and where we have used
\begin{equation}
	\int d\omega\, \omega [\tanh (\omega + a) - \tanh(\omega + b)]
	=
	b^2 - a^2.
\end{equation}

For small $T$, we have
\begin{equation}
\begin{aligned}
	\int_{-\infty}^{\infty} d\omega
	\,
	\omega
	\,
	\St_{\sfeb}[F]
	&\simeq
	-
	2\bar{g}^2
	\frac{\pi}{2}
	\int_{0}^{\infty} d\Omega
	\left\{
		\coth\left[\frac{\Omega}{2T}\right]
		-
		\coth\left[\frac{\Omega}{2\Te(\bar{x})}\right]
	\right\}
	\Omega^2
	\\
	&=
	-
	2\bar{g}^2
	\frac{\pi}{2}
	(2T)^3
	\int_{0}^{\infty} dy
	\,
	y^2
	\,
	\left\{
		\coth y
		-
		\,
		\coth\left[\frac{T}{\Te(x)} y\right]
	\right\}
	\\
	&=
	-
	2\bar{g}^2
	\frac{\pi}{2}
	(2T)^3
	\frac{\zeta(3)}{2}
	\left[
		1
		-
		\frac{\Te^3(\bar{x})}{T^3}
	\right]
	\\
	&=
	4\pi \zeta(3)
	\bar{g}^2
	\left[
	\Te^3(\bar{x})
	-
	T^3
	\right],
\end{aligned}
\end{equation}
In the long-wire limit, the left-hand side of Eq.~(\ref{eq:Te_ode}) 
can be dismissed and thus the effective temperature in the long wire limit is   
\begin{equation}
\label{eq:Te_small_T}
	T_e
	\simeq
	\left[
		\frac{
			E_{\mathsf{Th}}
		}{
			2\pi \zeta(3)
			\bar{g}^2
		}
		(e V)^2
		+
		T^3
	\right]^{1/3},
\end{equation}
which is \blue{Eq.~(3)} in the main text.


\section{Shot noise in the equilibrium boson regime}

Using \blue{Eqs.~(7), (8), (10), (17)} in the main text, we have
\begin{equation}
S_{\mathsf{shot}}
=
4
\,
\sigma_{\mathsf{D}}
\,
\int_0^1 d\xbar
\,
\Te(\xbar)
\,
\calJ(\xbar,\bar{g}^2, T)
,
\end{equation}
where
\begin{eqnarray}
\label{eq:calJ}
	\calJ(\xbar,\bar{g}^2, T)
	&=&
	\frac{1}{2}
	\int_{-\infty}^{\infty} 
	d \bar{\omega}
	\,
	\frac{
		\calI_{\mathsf{B}}(\xbar,\bar{\omega})
		+
		\calI_{\mathsf{MT}}(\xbar,\bar{\omega})
	}{
		1 - \im \Sigma^R_{2\Te \bar{\omega}}(\xbar)/\gammael
	}
,
	\\
	\calI_{\mathsf{B}}(\bar{\omega})
	&=&
	1 - \tanh^2 (\bar{\omega} - V \xbar/2\Te),
\end{eqnarray}
and
\begin{equation}\label{eq:IMTFinal}
	\calI_{\mathsf{MT}}(\bar{x},\bar{\omega})
	=
	-
	\frac{
		\bar{g}^2 \Te
	}{
		\gammael
	}
	\nint_{-\infty}^{\infty} d \bar{\nu}
	\,
	\tan^{-1}\left(\frac{2a \, \Te \bar{\nu}}{T}\right)
	\calF(2\Te \bar{\omega},2\Te \bar{\nu})
	.
\end{equation}
The function $\calF(\omega,\nu)$ is given by Eq.~(\ref{eq:MT_thermal_F}) and the parameter $a = \alpha/\alpha_m$. 
To obtain the MT contribution $\calI_{\mathsf{MT}}(\xbar,\bar{\omega})$, we performed the momentum integrals in a manner similar to Eq.~(\ref{eq:shot_MT_eqm}).

%

\subsubsection{$T = 0$ limit}

At $T = 0$, $F_{B,\nu} = \sgn(\nu)$. 
The frequency integral in the self-energy term \blue{[Eq.~(9)]} can be done exactly. 
The result is
\begin{equation}
\label{eq:selfE_T=0}
	\frac{1}{\gammael}
	\im \Sigma^R_{\omega}(x)
	\bigg|_{T = 0}
	=
	-
	\frac{
		\bar{g}^2
		\pi
		\,
		\Te(\xbar)
	}{
		\gammael
	}
	\,
	\ln 
	\left[
		2 \cosh (\bar{\omega} - V \xbar/2\Te)
	\right]
	.
\end{equation}
We then shift $\bar{\omega} \rightarrow \bar{\omega} + V\bar{x}/2\Te$ 
in Eq.~(\ref{eq:calJ})
and perform the dimensionless $\bar{\nu}$ integral in $\calI_{\mathsf{MT}}$ 
[Eq.~(\ref{eq:IMTFinal})]
to obtain
\begin{equation}
\label{eq:I_MT_T=0}
	\calI_{\mathsf{MT}}(\xbar, \bar{\omega})
	\bigg|_{T = 0}
	=
	-
	\frac{
		\bar{g}^2
		\pi
		\,
		\Te(\xbar)
	}{
		2\gammael
	}
	K(\bar{\omega}),
\end{equation}
\begin{equation}
\label{eq:Kw}
\begin{aligned}
	K(\bar{\omega})
	&\equiv
	\int_{-\infty}^{\infty} 
	d \bar{\nu}
	\,
	\begin{Bmatrix}
		-
		\begin{bmatrix}
			\tanh \bar{\omega} \, \tanh (\bar{\omega} + \bar{\nu}) + 1
			-
			\tanh^2 (\bar{\omega} + \bar{\nu})
			-
			\tanh^2 \bar{\omega}
		\end{bmatrix}
		\\
		+
		\sgn(\nu)
		\,
		\tanh \bar{\omega} \tanh (\bar{\omega} + \bar{\nu})
		\left[
			\tanh (\bar{\omega} + \bar{\nu})
			- 
			\tanh \bar{\omega}
		\right]
	\end{Bmatrix}
	\\
	&=
	-\sech^2 \bar{\omega}
	\,
	[
		2 + (\e^{2\bar{\omega}} - 1)\, \bar{\omega}
	]
	+
	2\tanh^2 \bar{\omega} 
	\,
	\ln (1 + \e^{2\bar{\omega}})
	.
\end{aligned}
\end{equation}
Combining Eqs.~(\ref{eq:calJ}), (\ref{eq:selfE_T=0}) and (\ref{eq:I_MT_T=0}), we have
\begin{equation}
\label{eq:shot_eqm_boson_T=0}
\shot
\bigg|_{T = 0}
=
4
\,
\sigma_{\mathsf{D}}
\int_0^1 d\xbar
\,
\Te(\xbar)
\,
\calJ_0
\left[
	\frac{
		\bar{g}^2
		\pi
		\,
		\Te(\xbar)
	}{
		\gammael
	}
\right]
,
\end{equation}
where the dimensionless function
\begin{equation}
\label{eq:calJ0}
	\calJ_0(y)
	=
	\int_{0}^{\infty} d \bar{\omega}
	\frac{
		(1 - \tanh^2 \bar{\omega})
		-
		\frac{y}{2}
		K(\bar{\omega})
	}{
		1 
		+
		y
		\,
		\ln 
		\left(
			2	\cosh \bar{\omega}
		\right)
	}
.
\end{equation}
Although $K(\bar{\omega})$ is complicated, it can be well approximated by 
\begin{equation}
	K(\bar{\omega}) 
	\simeq 
	- 2 \, \sech^2 \bar{\omega}.
\end{equation}
We then have
\begin{equation}
\begin{aligned}
	\calJ_0(y)
	&=
	\int_{0}^{\infty} d\bar{\omega}
	\frac{
		\sech^2 \bar{\omega} \,(1 + y)
	}{
		1 
		+
		y
		\,
		\ln 
		\left(
		2
		\cosh \bar{\omega}
		\right)
	}
	\\
	&\simeq
	\int_{0}^{\infty} d\bar{\omega}
	\frac{
		\sech^2 \bar{\omega} \, (1 + y)
	}{
		1 
		+
		y
		\,
		\ln 2
	}
	\\
	&=
	\frac{
		1 + y
	}{
		1 
		+
		y \,\ln 2
	}
	.
\end{aligned}
\end{equation}
In the long wire limit, $\Te$ is approximately a constant except at the proximity of the contacts. 
Thus, by using Eq.~(\ref{eq:Te_small_T}), we have at $T = 0$ that 
\begin{equation}
	S_{\mathsf{shot}}\bigg|_{T = 0}
	\simeq
	4\sigma_{\mathsf{D}}
	\,
	T_{\sfeff}
	\calJ_0
	\left(
		\frac{
			\bar{g}^2
			\pi
			\,
			T_{\sfeff}
		}{
			\gammael
		}
	\right),
\end{equation}
which corresponds to the Fano factor in \blue{Eq.~(5)} of the main text.


\section{Comparison of the effective electron temperature to the heavy-fermion experiment Ref.~\cite{shot_Chen_Natelson_arxiv_2023}}

\begin{figure}[hb!]
\centering
\includegraphics[width=0.8\linewidth]{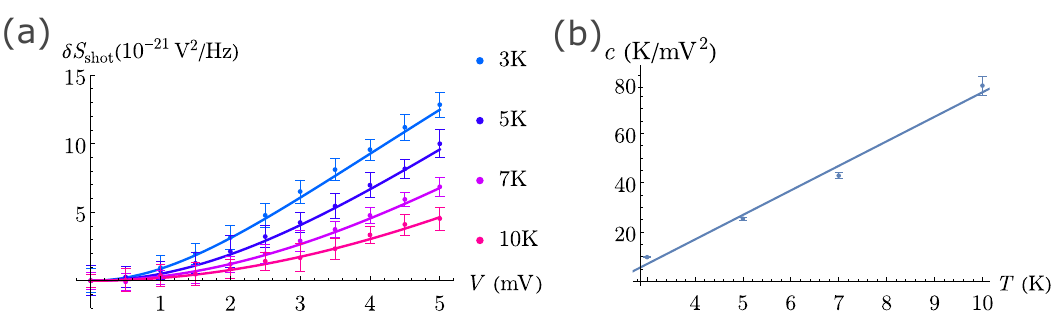}
\caption{
(a) Plot of the relative shot noise versus voltage at various temperatures. 
The filled dots represent experimental data from Ref.~\cite{shot_Chen_Natelson_arxiv_2023}
and the solid curves depict the fit based on Eq.~(\ref{eq:shotplot}). 
(b) Plot of the fitting parameter $c(T)$ as a function of temperature $T$. 
The solid line is the linear fit. 
}
\label{fig:fano}
\end{figure}

In this section, we compare the shot noise data for the YbRh$_2$Si$_2$ nanowire measured in the experiment Ref.~\cite{shot_Chen_Natelson_arxiv_2023} to
the results of our theory. 
Specifically, we fit the noise data using Eq.~(4) in the main text: 
\begin{equation}\label{eq:shotplot}
	\delta S_{\mathsf{shot}} 
	= 
	A 
	\left\{
		\left[
			T^4 + c(T) \, V^2
		\right]^{1/4} 
		- 
		T
	\right\},
\end{equation}
where $A$ and $c(T)$ are some parameters to be determined. 
Since the noise data were measured relative to the equilibrium one, we subtracted off the $V = 0$ part in the above equation.
We empirically set $A = 10^{-20} \mathrm{V^2/(Hz \cdot K)}$ and focus on the temperature dependence of $c(T)$. 
As shown in Fig.~\ref{fig:fano}(a), the shot noise data can be well-described by Eq.~(\ref{eq:shotplot}). 
The parameter $c(T)$ approximately varies with $T$ in a linear fashion [Fig.~\ref{fig:fano}(b)], as expected by our theory [Eq.~(4) in the main text]. 
\\\\
However, we note that YbRh$_2$Si$_2$ exhibits linear-$T$ resistivity while our model expects a constant Drude resistivity at the semiclassical level \cite{MFL_Wu_Liao_Foster_PRB_22}, which potentially accounts for the deviation from the linear-$T$ dependence in the parameter $c$. 
Moreover, this heavy-fermion material is strictly speaking three-dimensional, whereas our theory focuses on a 2D disordered marginal Fermi liquid. 
Whether the quantum critical fluctuations are effectively planar is still an open question. 
It would be interesting to explore the temperature dependence of shot noise in other (2D) quantum critical systems to further test our theory.
